\documentclass[12pt,preprint]{aastex}
\usepackage{epsf}
\usepackage{natbib}
\usepackage{hyperref}

\tabletypesize{\scriptsize}

\def\gappeq{\mathrel{ \rlap{\raise.5ex\hbox{$>$}}
                      {\lower.5ex\hbox{$\sim$}}  } }

\begin{document}
\slugcomment{Accepted for Publication in ApJ}
\shorttitle{Searching for Exozodis  with Kepler}
\shortauthors{Stark et al.}

\title{A Search for Exozodiacal Clouds with \emph{Kepler}}

\author{Christopher C. Stark\altaffilmark{1}, Alan P. Boss\altaffilmark{1}, Alycia J. Weinberger\altaffilmark{1}, Brian K. Jackson\altaffilmark{1}, Michael Endl\altaffilmark{2}, William D. Cochran\altaffilmark{2}, Marshall Johnson\altaffilmark{3}, Caroline Caldwell\altaffilmark{3}, Eric Agol\altaffilmark{4}, Eric B. Ford\altaffilmark{5}, Jennifer R. Hall\altaffilmark{6}, Khadeejah A. Ibrahim\altaffilmark{6}, Jie Li\altaffilmark{7}}

\altaffiltext{1}{Department of Terrestrial Magnetism, Carnegie Institution of Washington, 5241 Broad Branch Road, NW, Washington, DC 20015-1305; cstark@dtm.ciw.edu}

\altaffiltext{2}{McDonald Observatory, University of Texas at Austin, Austin, TX 78712}

\altaffiltext{3}{Department of Astronomy, University of Texas at Austin, Austin, TX 78712}

\altaffiltext{4}{Department of Astronomy, University of Washington, Seattle, WA 98195}

\altaffiltext{5}{Department of Astronomy, University of Florida, Gainesville, FL 32611}

\altaffiltext{6}{Orbital Sciences Corporation/NASA Ames Research Center, Moffett Field, CA 94035}

\altaffiltext{7}{SETI Institute/NASA Ames Research Center, Moffett Field, CA 94035}

\begin{abstract}
Planets embedded within dust disks may drive the formation of large scale clumpy dust structures by trapping dust into resonant orbits.  Detection and subsequent modeling of the dust structures would help constrain the mass and orbit of the planet and the disk architecture, give clues to the history of the planetary system, and provide a statistical estimate of disk asymmetry for future exoEarth-imaging missions.  Here we present the first search for these resonant structures in the inner regions of planetary systems by analyzing the light curves of hot Jupiter planetary candidates identified by the \emph{Kepler} mission.  We detect only one candidate disk structure associated with KOI 838.01 at the $3\sigma$ confidence level, but subsequent radial velocity measurements reveal that KOI 838.01 is a grazing eclipsing binary and the candidate disk structure is a false positive.  Using our null result, we place an upper limit on the frequency of dense exozodi structures created by hot Jupiters.  We find that at the 90\% confidence level, less than $21\%$ of \emph{Kepler} hot Jupiters create resonant dust clumps that lead and trail the planet by $\sim90^{\circ}$ with optical depths $\gtrsim5\times10^{-6}$, which corresponds to the resonant structure expected for a lone hot Jupiter perturbing a dynamically cold dust disk 50 times as dense as the zodiacal cloud.

\end{abstract}


\section{Introduction}
\label{intro}

Our inner solar system hosts a diffuse cloud of dust known as the zodiacal cloud.  These dust grains, originating from comets and asteroids, spiral in toward the Sun on timescales of millions of years due to a special relativistic Poynting-Robertson (PR) drag force, filling the inner few AU of the Solar System.  Nearly 23 years ago \citet{jz89} predicted that asteroidal dust would be trapped by Earth's gravity into exterior mean motion resonances, creating a clumpy ``resonant ring" of dust that rotates in lock with the Earth.  A half decade later similar models by \citet{d94} explained a puzzling asymmetry in the \emph{Infrared Astronomical Satellite}'s (\emph{IRAS}) observations of the zodiacal cloud: the thermal emission from grains trailing the Earth was brighter than that leading the Earth \citep{dnk88, r91}.  The hallmark of this model was the prediction of two dense clumps of dust leading and trailing the Earth, with the trailing clump more dense than the leading clump.  The following year, observations from the \emph{Cosmic Background Explorer} (\emph{COBE}) produced the first Earth-centric images of these dust clumps \citep{r95}.

With the exponential increase in the number of known exoplanets and the now-commonplace detection of dust disks around other stars \citep{mcm08}, it is not unreasonable to posit that extrasolar planetary systems harbor similar clumpy resonant rings in their ``exozodiacal" clouds.  Detection of these resonant rings would provide another parallel between our planetary system and others, and would inform our understanding of how planetary systems evolve, much in the same way the asteroid and Kuiper Belts reveal the history of our Solar System \citep[e.g.][]{clm07}.  Models show that the geometry of resonant rings strongly depends on the perturbing planet's mass and orbital parameters, as well as parameters that describe the dust disk \citep[e.g.][]{kh03, dm05, rba08, sk08}.  Detection of these rings would provide a new quantitative method to constrain the planet mass, dust grain size, disk geometry, and amount of dust in these systems.

Detection of resonant rings may also shed light on a potential problem for future missions that aim to directly image Earth-like extrasolar planets.  For a sufficiently small telescope, or sufficiently distant targets, resonant clumps may act as a source of confusion and mimic unresolved planetary companions.  Additionally, these clumpy structures may act as a source of noise and obfuscate the planet, if the system is close to edge-on \citep{ltu09}.  A limit on the frequency of occurrence of resonant rings with a given optical depth would prove useful for planning such a mission.

\begin{figure}
\begin{center}
\includegraphics[width=3in]{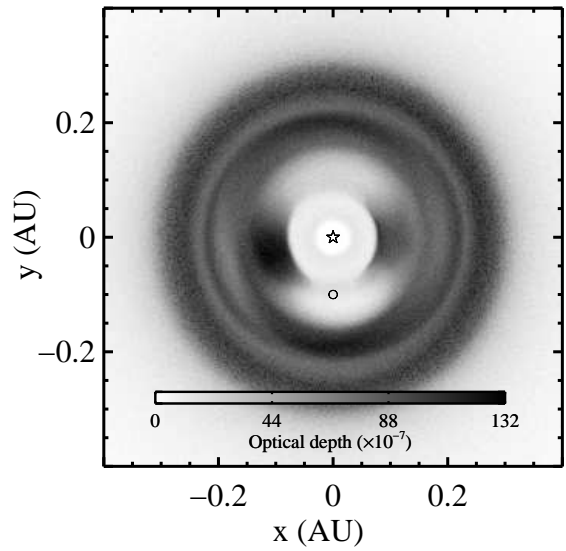}
\caption{Optical depth histogram of a resonant ring structure created by a Jupiter mass planet orbiting at 0.1 AU in a disk $\sim100$ times the density of our zodiacal cloud.  The planet, marked with a circle, orbits counter-clockwise and creates two distinct clumps of dust leading and trailing the planet by $\sim90^{\circ}$ in its orbit. \label{optical_depth_figure}}
\end{center}
\end{figure}

As was the case with the first detected exoplanets, we may expect to discover the extreme cases of these resonant ring structures first.  The models of \citet{s11} suggest that hot Jupiters embedded within disks hundreds of times more dense than our zodiacal cloud, and orbiting just outside of the circumstellar distance at which dust sublimates, can create extremely asymmetric resonant ring structures with two distinct clumps leading and trailing the planet by $\sim90^{\circ}$, as shown in Figure \ref{optical_depth_figure}.  If such a system were viewed edge-on, these dust clumps would pass in front of their host stars as they orbit in lock with the planet, temporarily dimming the star.  Unlike the planetary transit, the dust clumps would exhibit smooth, broad transit features, lasting for significant fractions of the planet's orbit.  \citet{s11} calculated the amplitude of this transit signal and predicted that such a signal may be currently detectable with \emph{Kepler}.

Here we present the first search for exozodiacal resonant dust structures within the \emph{Kepler} data.  In Section \ref{observations} we discuss the \emph{Kepler} data used for this search, our data reduction techniques, and our methods of signal detection.  In Section \ref{results} we present a single candidate exozodi structure associated with a \emph{Kepler} planet candidate, along with radial velocity (RV) measurements that show the candidate is a grazing eclipsing binary with twice the reported period and therefore a false positive.  We use our null result to place an upper limit on the frequency of dense exozodiacal structures.  Finally, in Sections \ref{discussion} and \ref{conclusions} we discuss our null result and provide a clear interpretation of our upper limit on exozodi structures.

\section{Data Reduction and Detection Methods}
\label{observations}

\subsection{Data Reduction}
\label{data_reduction}

We limited our search for exozodiacal ring structures to the list of $2\mathord{,}300$ Kepler Objects of Interest (KOIs) identified by \citet{b12}.  For planets on low eccentricity orbits, a resonant ring structure created by the planet revolves in lock with the planet's orbital motion.  As a result, the associated photometric variations are phase-locked with the planet and have a periodicity equal to the planet's orbit, making signal detection much easier for systems with identified planet candidates.

\citet{s11} predicted that Jupiter-mass planets create resonant rings with the largest degree of asymmetry.  We therefore selected only those KOIs larger than 8 Earth radii, a cut that should include the vast majority of Jupiter-mass planets given the mass-radius relationship of known planets \citep[e.g.][]{kg12}, and searched for minima in the light curve leading and trailing the planet by $90^{\circ}$.  We also required that these planets have blackbody equilibrium temperatures less than 1600 K, a rough upper limit on the sublimation temperature of silicate dust \citep[e.g.][]{dgt96,kmd09}.  Finally, we selected only KOIs with orbital periods $P < 20$ days because longer period KOIs lacked enough orbits to reliably estimate the uncertainty of the reduced, phase-folded light curve.  These selections resulted in a total of 106 KOIs used for our exozodi search.

For each of these candidates, we examined the first 11 quarters of \emph{Kepler} data (Q0 -- Q10), covering more than two years of observations.  The \emph{Kepler Mission} was designed to detect the short-lived transits of Earth-sized planets and has a photometric precision of $\sim2 \times 10^{-5}$ over 6.5 hr timescales for $V=12$ Sun-like stars \citep{jcc10, kbb10}.  Photometric variations caused by a resonant ring of dust can have an amplitude $\sim 10^{-4}$, but over a much longer timescale (on the order of days), during which \emph{Kepler} photometry fluctuates.  We therefore required many orbit foldings to build up a sufficient signal-to-noise ratio and obtain an accurate estimate of the light curve uncertainty.

To retain the low-amplitude, long-period resonant ring signals, we reduced the raw \emph{Kepler} data using the {\sc kepcotrend} routine from the {\sc pyke} software package\footnote{\href{http://keplergo.arc.nasa.gov/ContributedSoftware.shtml}{http://keplergo.arc.nasa.gov/ContributedSoftware.shtml}}\citep{bsj12}.  This routine removes systematic fluctuations from a \emph{Kepler} light curve by cotrending, i.e. removing photometric fluctuations correlated across many stars in the same CCD channel.  These correlations are recorded in cotrending basis vectors (CBVs) made available by the Kepler Science Office\footnote{CBVs are available at \href{{http://archive.stsci.edu/kepler/cbv.html}}{http://archive.stsci.edu/kepler/cbv.html}.  See the Kepler Data Characteristics Handbook for a fuller description of CBVs (\href{http://archive.stsci.edu/kepler/documents.html}{http://archive.stsci.edu/kepler/documents.html}). The version released 17 August 2011 contains the relevant description of the CBV beginning on p. 50.}.  For each quarter of data, there are 16 CBVs.  The optimal number of CBVs to use for cotrending is target-dependent.  Cotrending with more CBVs removes more fluctuations from a light curve, but using too many CBVs can distort real signals.  Generally speaking, cotrending with eight CBVs removes the majority of systematic trends.  Our analysis showed that six CBVs sufficiently removed systematic fluctuations without significantly impacting any potential exozodi signal, so we used six CBVs.

After cotrending the data, we normalized each quarter's data by its median, stitched them together, and removed all identified transits from the light curve.  To remove statistical outliers, we phase-folded the resulting light curve, binned the light curve into 30 bins, estimated the uncertainty within each bin from the standard deviation, and removed the data points more than four standard deviations from the median.  Generally speaking, the cotrended light curves of most KOIs with orbital periods less than 20 days exhibited normally distributed data within a bin, making the standard deviation a reliable uncertainty estimate.

We then temporarily filled all gaps in the data that were longer than twice the transit duration with placeholder values equal to the median of the light curve (no transit events were filled) and interpolated to an evenly spaced time base using a cubic spline in order not to introduce false periodic signals.  We Fourier-filtered each quarter separately, removing periodic fluctuations with timescales longer than $1.1P$ and shorter than $0.2P$.  We found these to be the tightest possible Fourier filter limits that did not significantly alter any of the models of \citet{s11}.  We then removed the placeholder values, as well as any data points that fell within 0.33 days of the endpoints of the placeholder gaps,  median-normalized each quarter, and stitched them back together to form a reduced light curve.

\begin{figure}
\begin{center}
\includegraphics[width=6.5in]{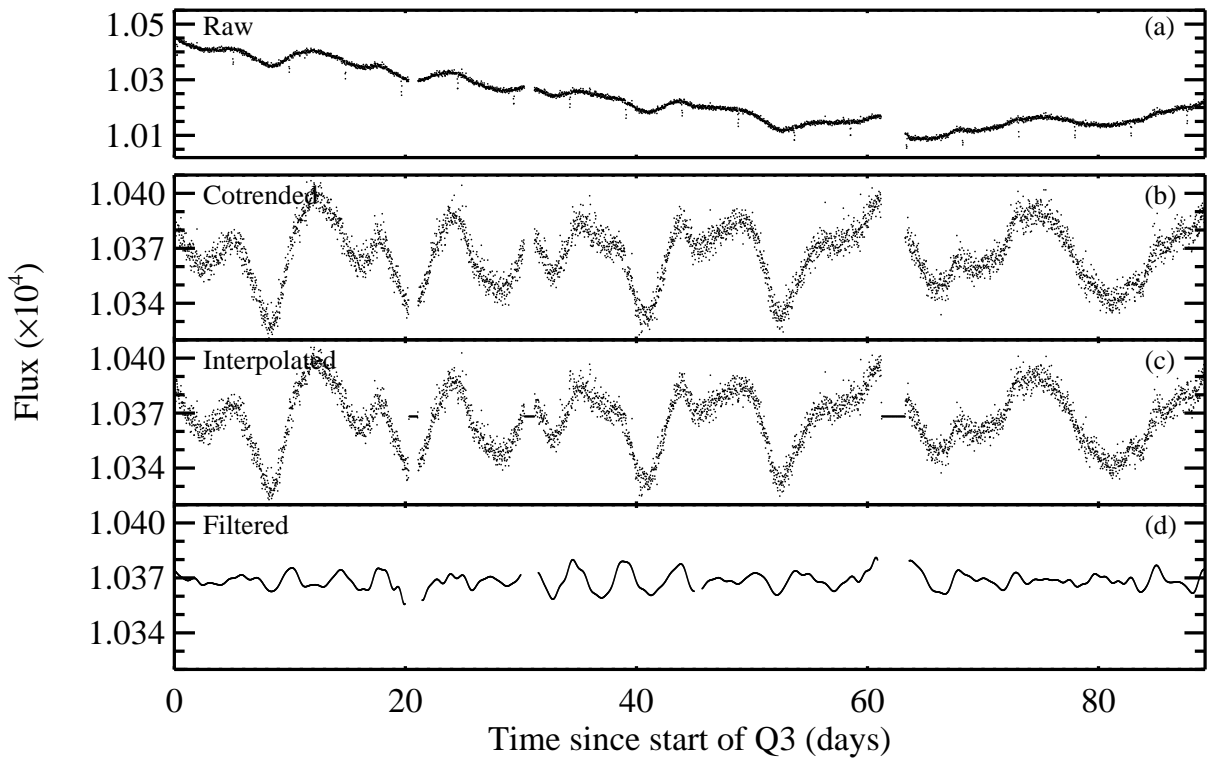}
\caption{Example of our data reduction process.  (a) The raw Q3 data for KOI 838.01.  (b) The data after cotrending with six CBVs and masking of planetary transits.  (c) The data after interpolating to an evenly spaced time grid and gap-filling with the median of the Q3 data.  (d) The data after Fourier filtering high and low frequencies and masking of points near gaps. \label{unfolded_lightcurve_figure}}
\end{center}
\end{figure}

Figure \ref{unfolded_lightcurve_figure} shows an example of our data reduction method applied to a single quarter of data.  From top to bottom, we plot a) the Q3 data for KOI 838.01 in its raw form, b) after cotrending with six CBVs and removing all transits, c) after gap-filling and interpolation, and d) after Fourier filtering the data and removing data from the gap-filled regions.  The jitter of the light curve is significantly reduced via this method, most notably for fluctuations on timescales longer than the period of the planet, which would not be averaged out by simple phase-folding in many cases.

Lastly, we phase-folded the reduced light curves and normalized them to unity.  The models of \citet{s11}, which ignore the planetary transits, predict that a light curve exhibiting an exozodi structure signal reaches a maximum at the time of the planet transit.  Our data reduction process removed and smoothly interpolated over the planet transits, so we normalized to unity using the first and last 4\% of the light curve, i.e. within and just exterior to the interpolated planet transit.  This anchors the phase-folded light curves to unity at a planet phase of zero, as desired.  We set the final uncertainty in each data point equal to the local standard deviation of the filtered data by binning the phase-folded, filtered data into 30 equally spaced bins.  These phase-folded, filtered light curves constitute the final data we analyzed.

\subsection{Exozodi Detection Method \label{detection_section}}

\begin{figure}
\begin{center}
\includegraphics[width=6.5in]{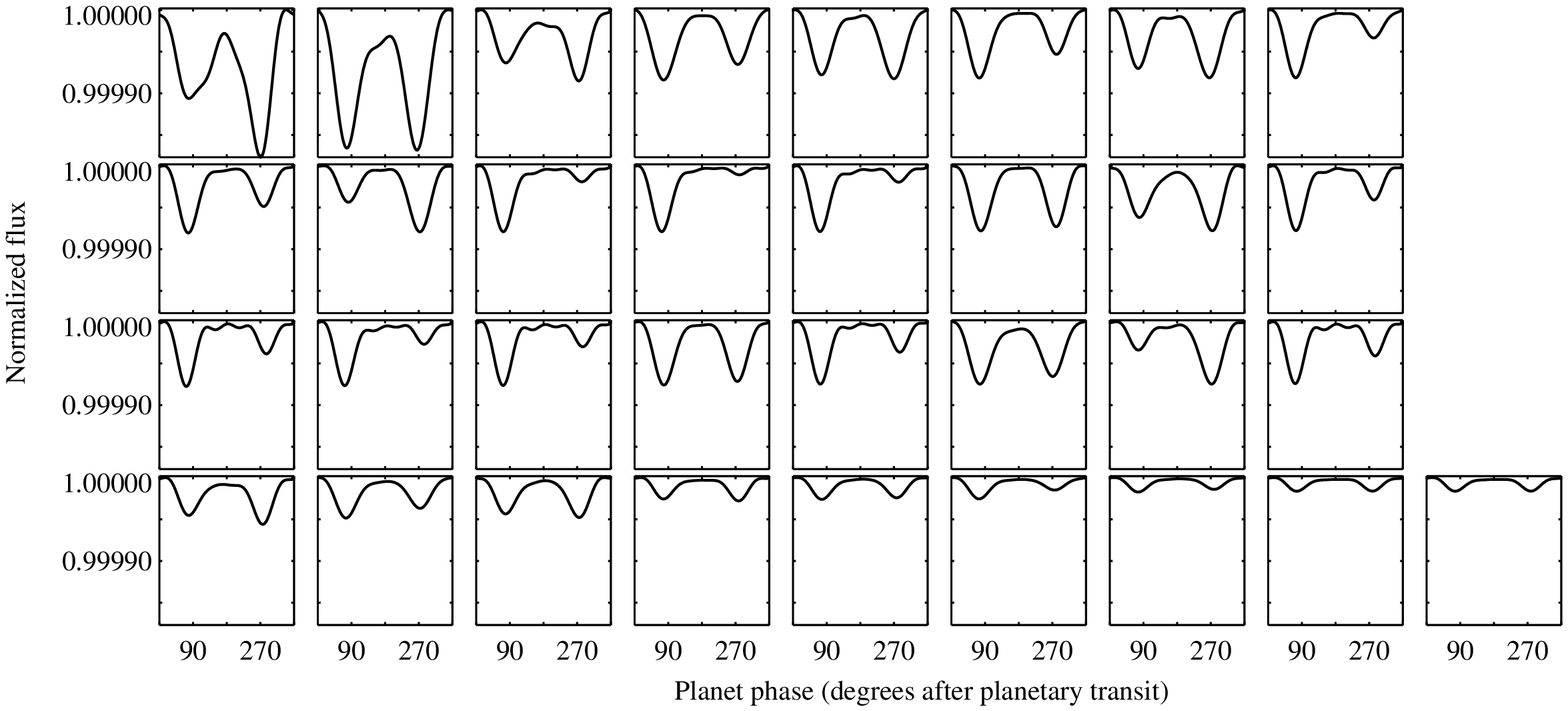}
\caption{Thirty-three model light curves used to fit the filtered, phase-folded light curve of each planet candidate.  During the fitting procedure, light curve amplitudes were allowed to vary from 0.5 to 5.5 times the amplitude shown.  \label{model_lightcurves_figure}}
\end{center}
\end{figure}

We searched for resonant ring signals by fitting each filtered, phase-folded light curve with the light curves of \citet{s11}.  We used all of the light curves predicted for exozodi structures created by Jupiter-mass planets on circular orbits around a Sun-like star, covering semi-major axes from 0.1 to 1 AU and disk optical depths from $10^{-6}$ to $5\times 10^{-5}$, i.e. 10 to 500 ``zodis."  We also included the non-collisional, single-grain size model constituents, i.e. the \citet{s11} ``seed models", and normalized their amplitudes to $8\times 10^{-5}$, roughly what would be expected for a 100 zodi disk.   In total, 33 model light curves were used to fit each planet candidate's filtered, phase-folded light curve.  Several of the model light curves exhibited a significant amount of noise, so we fit each of these light curves with a smooth interpolation to remove artifacts.  To interpolate the model light curves, we fit 10 sine and cosine terms to the light curve, using the first five integer multiples of the planet's period.  Figure \ref{model_lightcurves_figure} shows the 33 interpolated model light curves we used, ordered by the amplitude of the light curve.

The shape of a resonant structure's transit light curve depends on which of the planet's resonances are populated with dust.  Therefore the primary purpose of these models is to represent the possible resonances that inward-migrating dust grains can occupy when trapped by a Jupiter mass planet.  A Jupiter mass planet at 0.1 AU will trap dust grains of a given size primarily into a single resonance; a similar planet at 0.05 AU could trap dust of a slightly different size into the same resonance.  With this in mind, we fit the models to each KOI by matching the period of the models to that of the KOI.  We use these models only to detect the signal of a resonant structure, not to model the dynamics of any detected signal.

During the search process, we allowed the amplitude of each model to vary from 0.5 -- 5.5 times the modeled value and determined the best-fit model via $\chi^2$ minimization.  We then calculated the $\chi^2$ value for the best-fit model and the corresponding probability $Q$ that the reduced $\chi^2$ value would be larger by chance, where $Q = \Gamma(\nu/2,\chi^2/2)/\Gamma(\nu/2)$ and $\nu$ is the number of degrees of freedom.  A lower value of $Q$ indicates that the model has a greater chance of being wrong.  We also calculated $\chi_{\rm null}^2$ and the corresponding $Q_{\rm null}$, obtained by fitting the data with an unchanging light curve, i.e. a normalized flux equal to unity at all planet phases.

A candidate detection can then be defined by a light curve that exhibits a sufficiently large value of $Q$, greater than $\sim0.1$, and a sufficiently small value of $Q_{\rm null}$.  The exact values of $Q$ and $Q_{\rm null}$ that signify a detection are not well-defined without testing our detection sensitivity.  To determine the proper threshold values of $Q$ and $Q_{\rm null}$ that define a detection, we examined the false positive rate of a set of stars that are not known to harbor transiting planets.  We note that by comparing $Q$ values to those for a control sample, our detection method should be insensitive to any remaining weak correlations between data points that could reduce the effective value of $\nu$.

We selected a sample of $5\mathord{,}000$ stars that were not identified by \citet{b12} as planet candidate host stars to use as a control sample.  We do not expect to detect any real exozodi structures in the control sample because it is unlikely that a randomly selected control sample star would be oriented close  to edge-on and have both a massive planetary companion and dense exozodiacal disk.  We randomly drew these stars from the same set used for the Exoplanet Search Program and analyzed their first 11 quarters of data.  We selected only stars that had at least 8 quarters of data, such that the median number of quarters of data for our 106 KOIs (9.2 quarters) roughly equals the median number of quarters of data for our control sample (9.8 quarters).

\begin{figure}
\begin{center}
\includegraphics[width=6.5in]{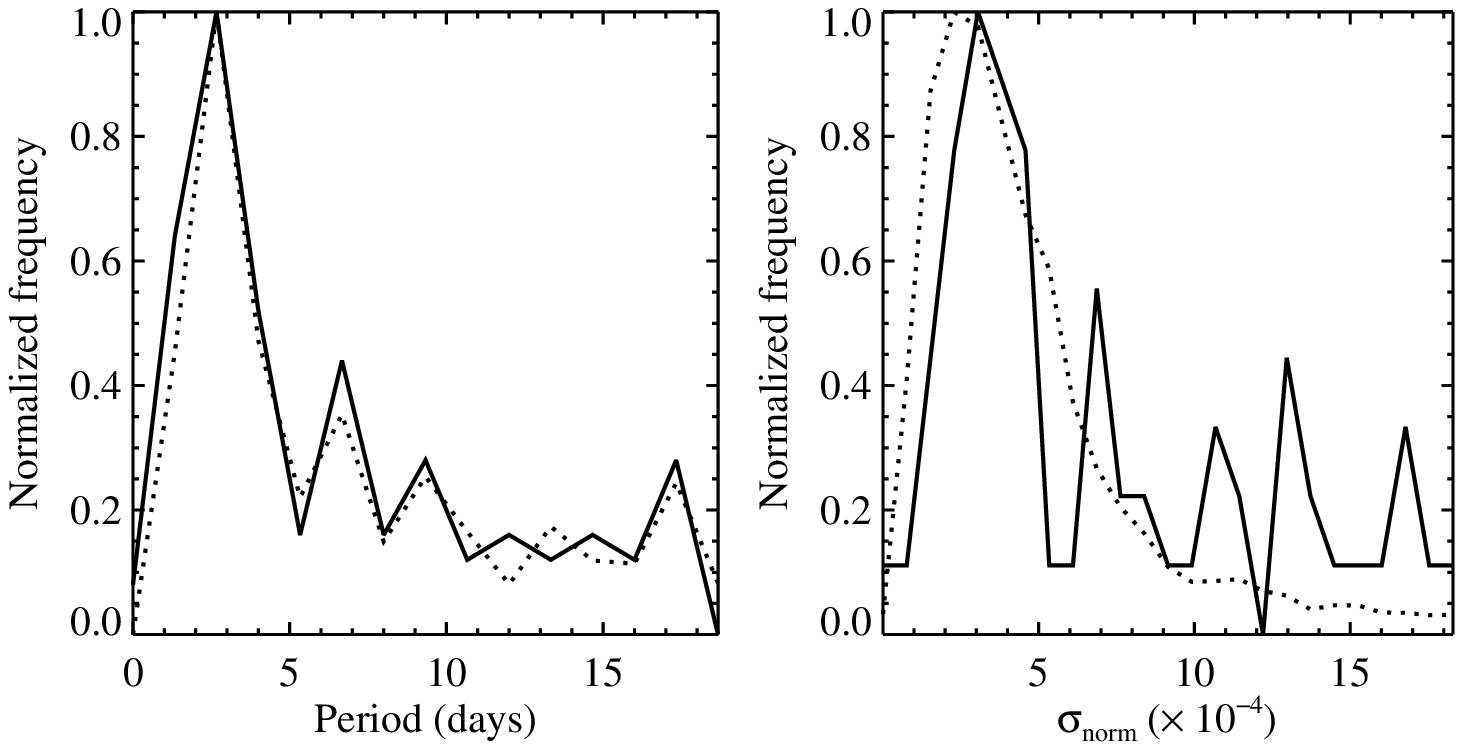}
\caption{\emph{Left:} Histogram of the orbital periods of all KOIs considered in this analysis (solid) and the non-planet candidates used as a control sample (dashed).  \emph{Right:} Histogram of the normalized stellar and photometric jitter of KOIs (solid) and non-planet candidates (dashed).  The peaks in KOI jitter for $\sigma_{\rm norm} >  6\times 10^{-4}$ are caused by small number statistics.  The control sample's period and jitter distributions closely match those of the KOIs.  \label{noise_and_period_figure}}
\end{center}
\end{figure}

We randomly assigned each of these stars 10 orbital periods for hypothetical planetary companions.  For each period, we assigned 10 hypothetical planetary phases, increasing the effective number of control sample objects to $500\mathord{,}000$.  We randomly drew the orbital periods of our hypothetical companions from the distribution of orbital periods of the 106 KOIs using a Monte Carlo method.  The 10 orbital phases were uniformly distributed between 0 and 1.  The left panel of Figure \ref{noise_and_period_figure} shows the distribution of orbital periods for the $500\mathord{,}000$ control sample objects and the 106 target KOIs; the period distributions match closely.

The right panel of Figure \ref{noise_and_period_figure} shows the median-normalized stellar and photometric jitter, $\sigma_{\rm norm}$, given by the standard deviation of the light curve divided by the median, for the control sample data and the KOI data after cotrending.  The peaks in KOI jitter for $\sigma_{\rm norm} >  6\times 10^{-4}$ are caused by small number statistics.  The distributions are similar; our control sample stars exhibit similar stellar and photometric jitter.

We reduced the data for our control sample in an identical fashion to that described in Section \ref{data_reduction}, but did not remove any hypothetical transits (there were none to remove) and only median-filled gaps longer than 4 hr.  We then fit the control sample (CS) light curves using the same exozodi model-fitting routine described above and recorded the values of $Q_{\rm CS}$ and $Q_{\rm CS,null}$ for each object.  Because all good fits within the control sample are by definition false positives, we defined the detection of a candidate exozodiacal resonant ring structure by the following process:
\begin{enumerate}
\item For each KOI, calculate $Q$ and $Q_{\rm null}$, and find the number of control sample objects with $Q_{\rm CS} > Q$ and $Q_{\rm CS,null} < Q_{\rm null}$.  This is the number of false positives $N_{\rm FP}(Q,Q_{\rm null})$.
\item Evaluate the binomial distribution function, 
\begin{equation}
\label{binomial_equation}
 P_B\left(x,n,p\right) = \frac{n!}{x! (n-x)!} p^x \left(1 - p \right)^{n-x},
\end{equation}
where $x = 0$ is the number of false positives allowed in our sample of $n=106$ KOIs, $p = N_{\rm FP}(Q,Q_{\rm null}) / N_{\rm CS}$ is the probability of detecting a false positive, and $N_{\rm CS}$ is the number of control sample stars.  $P_B$ is the confidence level of the detection; for a 3$\sigma$ detection, $P_B = 0.997$.
\end{enumerate}

\section{Results and Analysis}
\label{results}

\subsection{A False Positive Exozodi Associated with KOI 838.01}

Applying the detection methods described above, we found one exozodi candidate, KOI 838.01, which we detected at the 3$\sigma$ level.  \citet{b12} report that KOI 838.01 is a 12.3 Earth radius planet candidate on a 4.86 day orbit with an equilibrium blackbody temperature of 1036 K.  The \emph{Kepler} Input Catalog (KIC) lists the host star of KOI 838.01, KIC 5534814, as a G-type star with a radius of 0.991 $R_{\sun}$.

\begin{figure}
\begin{center}
\includegraphics[height=5in]{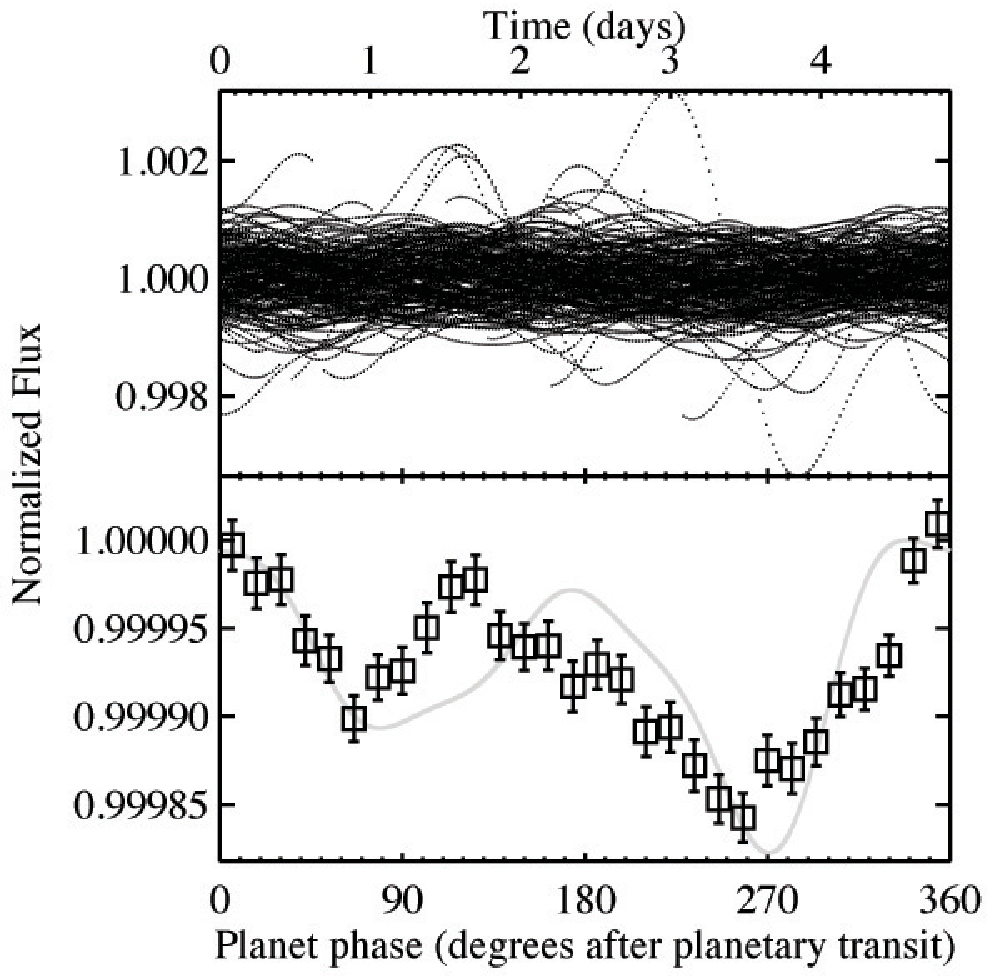}
\caption{\emph{Top:} Phase-folded light curve of KOI 838.01 after cotrending, interpolation, and filtering.  \emph{Bottom:} The same light curve, median-binned into 30 points with 1-$\sigma$ error bars.  The best-fit model used to detect the fluctuations is shown as a solid gray line.\label{folded_lightcurve_figure}}
\end{center}
\end{figure}

The phase-folded, filtered light curve of KOI 838.01, produced using all available data (Q1 - Q11), is shown in Figure \ref{folded_lightcurve_figure}.  The top panel shows all data points in the light curve; the periodic fluctuations are not obvious.  The bottom bins the data into 30 equally spaced bins and clearly shows the periodic fluctuations.  The uncertainty of each bin in the bottom panel was set equal to the standard deviation of the data within the bin divided by the square root of the number of data points in the bin.  The best fit model used to detect the light curve fluctuations is shown as a gray line, and corresponds to a \citet{s11} model of a Jupiter mass planet orbiting at 0.5 AU in a disk 500 times as dense as our zodiacal cloud.  This best fit corresponded to $Q = 0.25$, while a flat light curve gave $Q_{\rm null} = 4.8\times10^{-7}$.

To detect KOI 838.01's candidate exozodi structure, we considered only objects that had similar $\sigma_{\rm norm}$ ($0.5 < \sigma_{\rm norm} / \sigma_{\rm KOI\; 838.01, norm} < 1.5$) and similar periods ($0.5 < P / P_{\rm KOI\; 838.01} < 1.5$) to that of KOI 838.01.   This cut yielded 7 KOIs and $20\mathord{,}946$ control sample objects.  We detected only 6 false positives within these $20\mathord{,}946$ control sample stars.  Using the binomial distribution function above with $n=7$ and $p=6/20\mathord{,}946$, we calculated a detection confidence level of $99.8\%$.

We note that our detection criteria are relatively conservative.  We did not apply any criteria as to the amplitude of the signal and did not require that any of the false positives exhibit two distinct minima, as KOI 838.01 clearly does.  The addition of these criteria would only improve our detection statistics.  For example, requiring that the false positives exhibit a binned light curve amplitude greater than or equal to the amplitude of the KOI 838.01 binned light curve decreases the number of false positives to 3, increasing the detection confidence level to $99.9\%$.

\citet{b12} flag KOI 838.01 as having a V-shaped transit.  Candidates with V-shaped transits are likely to be diluted eclipsing binaries \citep{b12}, but we found no other evidence for binarity.  KOI 838.01 is not listed in the Kepler Eclipsing Binary Catalog\footnote{http://archive.stsci.edu/kepler/eclipsing\_binaries.html} \citep{spw11} and was not identified as an eclipsing binary by \citet{od12}.  For the reported orbital period there are no signs of a secondary transit in the \emph{Kepler} light curve, which we ruled out to a level of $10^{-4}$.  We also performed an odd/even analysis of KOI 838.01.  We calculated the depth of the odd and even transits using the median of a 30 minute wide bin centered on the transit minima and found odd and even transit depths of 6310$\pm$60 ppm and 6340$\pm$70 ppm, respectively; the transit depths are equal to within uncertainties and there is no evidence for an odd/even asymmetry.

In an attempt to rule out an eclipsing binary scenario for KOI 838.01, we observed its host star KIC 5534814 with the Tull Coude Spectrograph \citep{t95} at the Harlan J. Smith 2.7 m Telescope (HJST) at McDonald Observatory.  The Tull spectrograph covers the entire optical spectrum at a resolving power of $R=60,000$. We observed KIC 5534814 four times in 2012: July 29 and 30, August 7, and September 5 (UT).  At each visit we took either three or four 1200 second exposures that we co-added to one exposure.  

In conjunction to the KIC 5534814 observations, we always observed the \emph{Kepler} field RV standard star HD 182488. After bias-subtraction, flat-fielding, cosmic-hits removal, and extraction of the Echelle orders with standard IRAF routines, we cross-correlated the KIC 5534814 spectra with the standard star template to obtain the absolute RV of the target.  For this purpose we used the wavelength range from 4950 to 6800 \AA.

\begin{figure}
\begin{center}
\includegraphics[width=4in,angle=270]{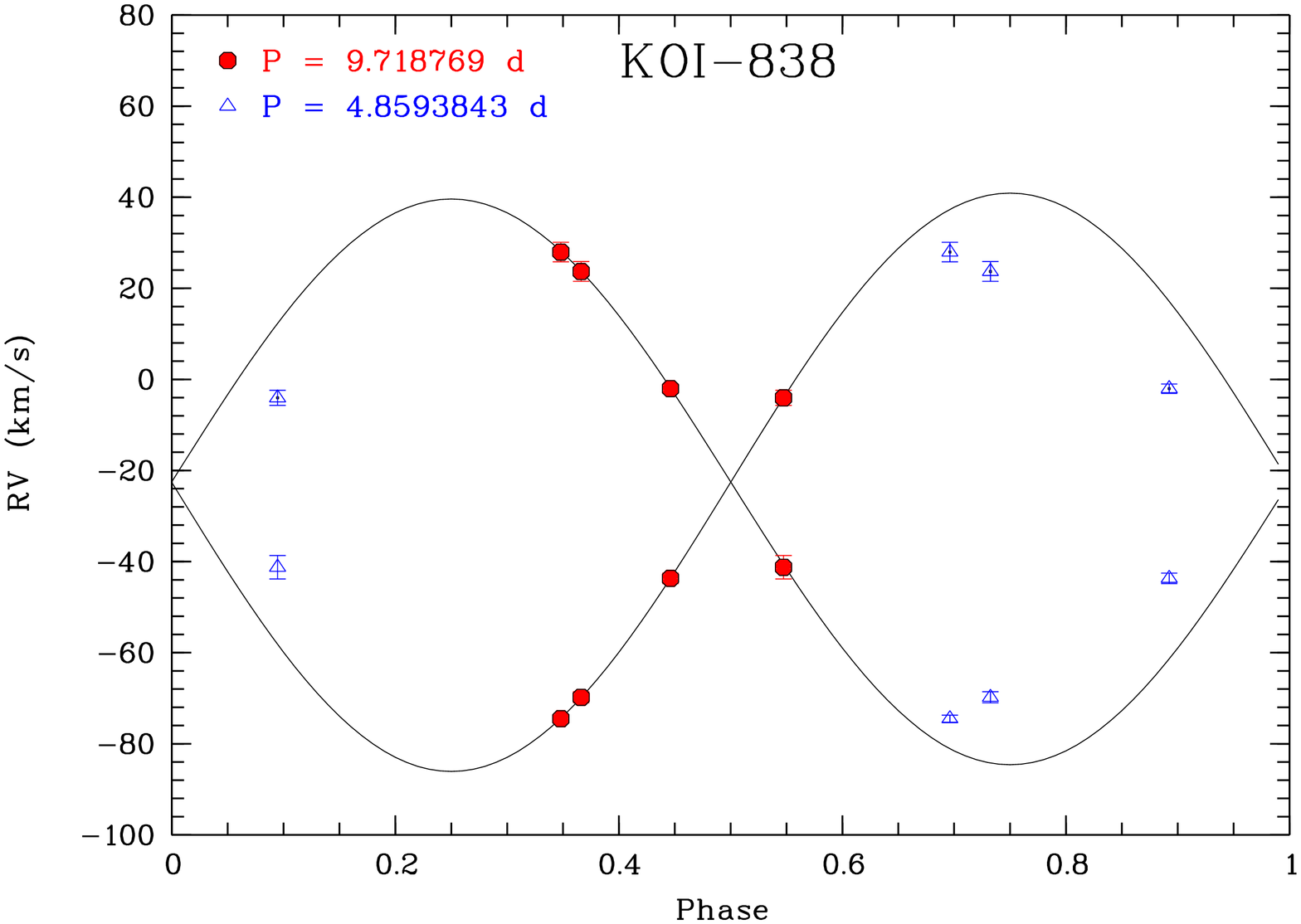}
\caption{RV measurements for KIC 5534814 phased to the period of the suspected planetary transit $P=4.86$~days (triangles) and to twice this period, $P=9.72$~days (filled circles). While the shorter period is not consistent with a circular orbit, the longer period is a good match to this scenario. The solid line represents the best-fit Keplerian RV orbit (assuming $e=0$ and adopting the period and phase from the photometry). This result suggests that KOI 838.01 is a nearly equal mass grazing eclipsing binary.  \label{rvfig}}
\end{center}
\end{figure}

We found the cross-correlation function of KIC 5534814 to be clearly double-peaked, indicating a double-lined spectroscopic binary (SB2).  As both cross-correlation peaks are of similar strength, we could not decide which of the two stars produced which peak at a given phase. However, the RVs phased up to twice the period of transit match a circular SB2 orbit very well (see Figure \ref{rvfig}). On the other hand, phasing the data to the original period does not match a circular orbit.  We thus conclude that KOI 838.01 is not a planetary companion, the system is a grazing eclipsing binary with a period of 9.719 days, and the ``exozodi" signal is a false positive.

We adopted this period and phase value, fixed the orbital eccentricity to 0, and determined RV semi-amplitudes of $K_{\rm A} = 63.5\pm0.14$\,km\,s$^{-1}$ and $K_{\rm B}=62.1\pm0.17$\,km\,s$^{-1}$.  The two stellar components are nearly equal in mass, with a mass ratio of $\approx 0.98$.  Assuming a circular orbit and equal stellar radii, our odd/even transit analysis suggests that the luminosity of the two stellar components are also nearly equal to within a few percent.  

We re-analyzed the phase-folded light curve of KOI 838.01 using the updated period of 9.719 days, masking off both primary and secondary transits.  The updated period reduces the minimum and maximum frequency of the fast Fourier transform (FFT) filters by a factor of two, and this new FFT bandpass allows through significantly more noise at smaller frequencies that was previously filtered.  As a result, the previously detected signal falls below the detection threshold.  Using the updated period of 9.719 days for KOI 838.01, our detection method produced no false positives.

\subsection{Constraining the Frequency of Exozodiacal Cloud Structures \label{exozodi_limits_section}}

With exception to the false positive KOI 838.01, we found no other signs of exozodiacal resonant ring structures at the 3$\sigma$ level in our sample of 106 \emph{Kepler} planet candidates with radii greater than 8 Earth radii, periods less than 20 days, and equilibrium temperatures less than 1600 K (for convenience, we refer to these as ``hot Jupiters").  We used this null result to place an upper limit on the frequency $f$ of clumpy resonant dust structures created by these hot Jupiters in exozodiacal clouds.  

To place an upper limit on $f$, we needed to know the detection efficiency of our methods.  To calculate this, we added several of the exozodiacal photometric variations modeled by \citet{s11} to our $5\mathord{,}000$ control sample stars to create simulated (SIM) light curves and attempted to detect them.  We used the models of a Jupiter mass planet at 0.5 AU in 10, 20, 50, and 100 zodi dust clouds.  We randomly drew orbital periods for the models from the distribution of planet candidate orbital periods shown in Figure \ref{noise_and_period_figure} and set the phase of all models to zero.  We normalized the model light curves to unity flux at a planet phase of zero and then multiplied the raw control sample light curves by the normalized model light curves to obtain the simulated light curves.

We then reduced the simulated light curves in a fashion identical to the control sample light curves and fit the phase-folded simulated light curves with our set of exozodi models,  recording the values of $\sigma_{\rm SIM,norm}$, $Q_{\rm SIM}$, and $Q_{\rm SIM,null}$.  However, to avoid any biases resulting from prior knowledge of the simulated light curves, we excluded the specific models that were incorporated into the simulated light curves during fitting.  We calculated the detection efficiency by dividing the number of simulated light curves that met some detection threshold by the total number of simulated light curves.

To place an upper limit on the frequency of exozodiacal cloud structures, we followed the procedure below:
\begin{enumerate}
\item Choose a value of $\sigma_{\rm 0}$ to serve as an upper limit on the acceptable normalized photometric jitter of a Fourier-filtered light curve, and choose values $Q_{\rm 0}$ and $Q_{\rm null,0}$ to set the detection threshold criteria for the entire sample.
\item Count the number of ``quiet" planet candidate light curves $N_{\rm PC}(\sigma_{\rm 0})$, ``quiet" control sample light curves $N_{\rm CS}(\sigma_{\rm 0})$, and ``quiet" simulated light curves $N_{\rm SIM}(\sigma_{\rm 0})$ that have normalized, filtered photometric jitter $<\sigma_{\rm 0}$.
\item Calculate the number of ``quiet" planet light curves $N_{\rm P}(\sigma_{\rm 0})$ = $N_{\rm PC}(\sigma_{\rm 0}) (1-{\rm FPR})$, where FPR is the fraction of \emph{Kepler} hot Jupiter candidates that are false positive planet candidates.  We adopt $\rm{FPR}=0.35$ for \emph{Kepler} hot Jupiter candidates \citep{sdm12}
\item Calculate the number of false positive exozodis in the quiet control sample $N_{\rm CS}^{\rm FP}(\sigma_{\rm 0},Q_{\rm 0},Q_{\rm null,0})$ that satisfy $Q_{\rm CS} > Q_0$ and $Q_{\rm CS,null} < Q_{\rm null,0}$.
\item Calculate the \emph{allowable} number of false positive exozodis in the quiet control sample $N_{\rm CS}^{\rm FP, allowed}(\sigma_{\rm 0})$ at the 3$\sigma$ level using the binomial distribution function given by Equation \ref{binomial_equation}, with $P_B = 0.997$, $n = N_{\rm P}(\sigma_{\rm 0})$, $x = 0$, and $p = N_{\rm CS}^{\rm FP, allowed}(\sigma_{\rm 0}) / N_{\rm CS}(\sigma_{\rm 0})$.  If $N_{\rm CS}^{\rm FP} < N_{\rm CS}^{\rm FP,allowed}$, the detection criteria in Step 1 meet the 3$\sigma$ threshold and we proceed.
\item Calculate the detection efficiency $E(\sigma_{\rm 0},Q_{\rm 0},Q_{\rm null,0})$ by dividing the number of quiet simulated light curves that meet the detection criteria $N_{\rm SIM}(\sigma_{\rm 0},Q_{\rm 0},Q_{\rm null,0})$ by the total number of quiet simulated light curves $N_{\rm SIM}(\sigma_{\rm 0})$.
\item Using the binomial distribution function, calculate the maximum frequency of resonant ring structure $f$ that would still result in 0 detections in our sample of quiet planets.  To do this, use Equation \ref{binomial_equation} with $x = 0$, $n = N_{\rm P}(\sigma_{\rm 0})$, $p = f E(\sigma_{\rm 0},Q_{\rm 0},Q_{\rm null,0})$, and $P_B = 1-C$, where $C$ is the confidence level of this limit.  We set $C = 0.9$.
\item Repeat the process for different values of $\sigma_{\rm 0}$, $Q_{\rm 0}$, and $Q_{\rm null,0}$, seeking to minimize $f$.
\end{enumerate}

We used the above procedure to calculate upper limits on the frequency of resonant exozodi structures at the 90\% confidence level.  Table \ref{exozodi_limits_table} lists these upper limits along with the optimal detection thresholds and associated detection efficiencies.  Because the detection efficiency increases with exozodi density, we can state that, at the 90\% confidence level, less than $56\%$ of \emph{Kepler} hot Jupiters produce exozodiacal dust structures with density asymmetries greater than those predicted by the 20 zodi model of \citet{s11}, $<21\%$ produce structures with asymmetries greater than predicted by the 50 zodi model, and $<18\%$ produce structures with asymmetries greater than predicted by the 100 zodi model.  We were unable to constrain the frequency of structures using the 10 zodi \citet{s11} model.

\begin{deluxetable}{cccccc}
\tablewidth{0pt}
\footnotesize
\tablecaption{Upper Limits on the Frequency of Resonant Exozodiacal Cloud Structures for \emph{Kepler} Hot Jupiters at the 90\% Confidence Level\label{exozodi_limits_table}}
\tablehead{
\colhead{Disk Density} & \colhead{$\sigma_{\rm 0}$} & \colhead{$Q_{\rm 0}$}  & \colhead{$Q_{\rm null,0}$} & \colhead{Detection Efficiency} & \colhead{Frequency Upper Limit} \\
\colhead{(zodis)\tablenotemark{*}}
}
\startdata
10   & $1.55\times10^{-4}$ & 0.263 & $5.83\times10^{-8}$ & 0.05 & Unconstrained\\
20   & $1.55\times10^{-4}$ & 0.263 & $5.83\times10^{-8}$ & 0.16 & 0.56\\
50   & $1.55\times10^{-4}$ & 0.263 & $5.83\times10^{-8}$ & 0.43 & 0.21\\
100 & $1.55\times10^{-4}$ & 0.263 & $5.83\times10^{-8}$ & 0.49 & 0.18\\
\enddata
\vspace{-0.1in}
\tablenotetext{*}{As defined by \citet{s11}}
\end{deluxetable}

\section{Discussion}
\label{discussion}

So far we have placed constraints on the asymmetry of exozodiacal resonant ring structures created by hot Jupiters by citing the \citet{s11} models.  It is useful to examine the geometry of these models to develop a simpler, more intuitive constraint in the context of a face-on disk.  A sample resonant ring structure created by a hot Jupiter in a 100 zodi disk is shown in Figure \ref{optical_depth_figure}.  The two dust ``clumps" that contribute to the exozodi transit features are clearly identifiable and have a characteristic size $\sim a$, where $a$ is the semi-major axis of the planet's orbit.  For the 20, 50, and 100 zodi models of \citet{s11}, the resonant dust clumps have face-on optical depths $\sim2\times10^{-6}$, $\sim5\times10^{-6}$, and $\sim10^{-5}$, respectively.  Therefore, we can state that at the 90\% confidence level, $<56\%$ of \emph{Kepler} hot Jupiters create resonant dust clumps that lead and trail the planet by $\sim90^{\circ}$ with optical depths $\gtrsim2\times10^{-6}$, $<21\%$ create such dust clumps with optical depths $\gtrsim5\times10^{-6}$, and $<18\%$ create such dust clumps with optical depths $\gtrsim10^{-5}$.

The limits on the frequency of dense resonant ring structures in exozodiacal clouds listed in Table \ref{exozodi_limits_table} are given with respect to the 106 hot Jupiter \emph{Kepler} planet candidates with blackbody equilibrium temperatures less than 1600 K and orbital periods less than 20 days.  Calculating the exozodi frequency limits with respect to all hot Jupiters with orbital periods less than 20 days, including those which orbit at distances interior to the dust sublimation radius, marginally reduces the frequency limits to 0.52, 0.19, and 0.17 for 20, 50, and 100 zodi disk structures, respectively.

Extrapolating these results to more distant Jupiter mass planets, perhaps near the habitable zone or beyond, is not straightforward.  Planets on orbits with larger semi-major axes have increased resonant trapping efficiencies, acting to enhance the dust clumps \citep[e.g.][]{sk08}.  However, this should be balanced in part by the fact that more distant planets typically have larger orbital eccentricities \citep[e.g.][]{kcg12}, which reduces trapping efficiency \citep[e.g.][]{dm05}.

The models of \citet{s11} placed the dust producing planetesimals just exterior to the planet's orbit.  Given this scenario, we might expect a higher disk frequency, and therefore frequency of resonant ring structures, for hot Jupiters on larger orbits where the disk lifetimes are longer.  However, dust originating from distant planetesimals can migrate inward over great distances via PR drag, even for dense 100 zodi disks; planets on compact orbits need not receive dust from young, compact planetesimal distributions.

Finally, and most significantly, hot Jupiters likely migrated inward to their current orbits, so their dynamical history differs from that of their more distant counterparts.  \citet{srf12} showed that \emph{Kepler} hot Jupiter planet candidates orbiting solar type stars show far fewer signs of transit timing variations than hot Neptunes, suggesting that stars with hot Jupiters lack additional planetary companions.  It's unclear whether this means that systems with hot Jupiters also lack the planetesimals needed to generate a debris disk.

For this analysis, we assumed that all \emph{Kepler} hot Jupiter candidates are massive enough to create resonant ring structures with density asymmetries similar to those of the Jupiter mass models of \citet{s11}.  Given that Neptune mass planets can create exozodi transit signatures that have magnitudes within a factor of 2 of their Jupiter mass counterparts, this was not an unreasonable assumption \citep{s11}.

Although we found no evidence for a resonant ring of dust associated with any of the \emph{Kepler} hot Jupiter candidates, many of these candidates' phase-folded light curves showed significant fluctuations.  We found 51 of our 106 hot Jupiter candidates' light curves (48\%) had $Q_{\rm null} < 10^{-6}$ and were inconsistent with being flat.  For comparison, 20\% of our control sample objects had $Q_{\rm null} < 10^{-6}$, i.e. 20\% of stars exhibit noise/oscillations that survive our filtering process to produce fluctuations in their phase-folded light curves.  The difference between these rates (28\%) suggests that $\sim30$ of the hot Jupiter candidates' light curves exhibit fluctuations somehow related to the planetary candidate.  Given that approximately 35\% of \emph{Kepler} hot Jupiter candidates are in fact eclipsing binaries \citep{sdm12}, many of these fluctuations may be a combination of ellipsoidal star variations, beaming, and reflection/emission commonly associated with eclipsing binaries (by eye, we estimate that this may be the case for up to 20 of the candidates).  Alternatively, the frequency of stellar oscillations among binaries may be linked to the period of the binary.

Our analysis detected one false positive, KOI 838.01, when phase folding the light curve of KOI 838.01 to the incorrect period reported in the literature.  This raises an underlying issue when searching for slowly varying signals within the \emph{Kepler} data set.  All light curves exhibit different noise properties (even different quarters within the same light curve) and so we are limited to detection via a statistical analysis.  We present the details of our best effort thus far at detecting the slow varying signal associated with a resonant dust disk in Section \ref{detection_section}.  Clearly we must rely on additional data to determine the true nature of any candidate exozodiacal dust signal, as we did for KOI 838.01 using radial velocity follow-up measurements.  Further, we note that a statistical detection method alone cannot differentiate between signals created by resonant dust and any other unknown phenomena that produce similar signals.  In the case of the binary KOI 838.01, the unexplained detected signal may be real and somehow related to the orbit of the binary.

Previous efforts to detect warm dust disks around \emph{Kepler} target stars with the \emph{Wide-field Infrared Survey Explorer} (\emph{WISE}) have produced few candidates \citep{krf11, rma12, lg12, kw12}.  As summarized by \citet{kw12}, many of these candidate detections have been revised to upper limits, and all but a few of the detections are statistically consistent with the chance alignment of background galaxies.  This low detection rate is in large part because \emph{WISE} is only sensitive to hot dust with fractional luminosities $\gtrsim10^{-3}$, or $\sim10\mathord{,}000$ zodis, around Sun-like stars \citep{kw12}.  Our disk detection method is nearly 3 orders of magnitude more sensitive than the \emph{WISE} searches, probing the \emph{Kepler} target stars for disks with densities down to 20 zodis, though not all 20 zodi disks will harbor detectable resonant dust structures.

Additional data from the extended \emph{Kepler} mission will reduce noise in the phase-folded light curves.  We expect the number of false positives to decrease, the detection efficiency to increase, and the limits on exozodi structure frequency to improve.  As an example, increasing the detection efficiency alone from 0.43 to 0.6 would reduce the upper limit on the frequency of resonant rings in disks with density $\gtrsim50$ zodis from 0.21 to 0.15.

In this first effort to photometrically detect the density asymmetries associated with planet-induced resonant ring structures in exozodiacal clouds, we have focused on the most extreme case of a Jupiter mass planet.  Future efforts to look for resonant ring structures created by Neptune mass planets may be more successful.  Neptune mass planets can create exozodi transit signatures with similar magnitudes \citep{s11} and are significantly more abundant.  \citet{b12} list 683 \emph{Kepler} planet candidates with radii between 2 and 8 Earth radii, equilibrium temperatures $<1600$ K, and periods less than 20 days.  Assuming a similar FPR to that of the Jupiter mass planets, $\sim0.3$, and a detection efficiency half that of the Jupiter mass planet case, we expect that a null result for the detection of resonant rings created by hot Neptunes in the current data set would limit the frequency of resonant structures in 100 zodi disk to less than 10\%.

\section{Conclusions}
\label{conclusions}

We used the first 11 quarters of \emph{Kepler} data to search for signs of exozodiacal resonant ring structures among \emph{Kepler} planet candidates identified by \citet{b12}.  We examined 106 planet candidates that were larger than 8 Earth radii, had a blackbody temperature less than 1600 K, and an orbital period less than 20 days.  We detected one candidate disk structure at the $3\sigma$ confidence level associated with KOI 838.01, but our follow-up RV observations showed a double-peaked cross-correlation function, indicating that KOI 838.01 is a grazing eclipsing binary and the exozodi signal is a false positive.  We found that the KOI 838.01 grazing eclipsing binary is best fit by two nearly equal mass stars on circular orbits with an orbital period of 9.719 days, twice the orbital period reported by \citet{b12}.

Using our null result, we placed an upper limit on the frequency of dense exozodi structures created by Jupiter mass planets.  At the 90\% confidence level, we find that $<56\%$ of \emph{Kepler} hot Jupiters are accompanied by resonant dust clumps that lead and trail the planet by $\sim90^{\circ}$ with optical depths $\gtrsim2\times10^{-6}$, $<21\%$ have such dust clumps with optical depths $\gtrsim5\times10^{-6}$, and $<18\%$ have such dust clumps with optical depths $\gtrsim10^{-5}$.  We expect that a future search for resonant ring structures created by the more abundant Neptune mass planets may have a higher chance of success, and a null result may limit the frequency of dense dust clumps created by hot Neptunes to $\lesssim10\%$.

\acknowledgments

The authors thank Guillem Anglada-Escud\'e and Evgenya Shkolnik for helpful discussions.  This work was supported by the Carnegie Institution of Washington.

\end{document}